\documentclass{PoS}

\usepackage{graphicx}
\usepackage{dcolumn}
\usepackage{bm}
\usepackage{amsmath}

\newcommand{\beq}{\begin{eqnarray}}
\newcommand{\eeq}{\end{eqnarray}}

\newcommand{\real}{{\sf I}\kern-.12em{\sf R}}
\newcommand{\comp}{{\sf I}\kern-.50em{\sf C}}
\newcommand{\unity}{{\sf I}\kern-.54em{\sf 1}}

\newcommand{\tef}{\theta_{\rm eff}}
\newcommand{\tief}{\theta_{I \rm eff}}
\newcommand{\esb}{\vec E \cdot \vec B}
\newcommand{\eisb}{\vec E_I~\cdot~\vec B}
\newcommand{\ceb}{\chi_{CP}\,}
\newcommand{\avq}{\langle Q \rangle}
\newcommand{\avqs}{\langle Q^2 \rangle}

\title{Effective $\theta$ term by CP-odd electromagnetic background fields}

\ShortTitle{Effective $\theta$ term by electromagnetic background $\esb$}

\author{Claudio Bonati\\
        Dipartimento di Fisica, Universit\`a di Pisa and INFN, Largo P
ontecorvo 2, I-56127 Pisa, Italy\\
        E-mail: \email{bonati@df.unipi.it}}
        
\author{Guido Cossu\\
        High Energy Accelerator Research Organization (KEK), Ibar
aki 305-0801, Japan\\
        E-mail: \email{cossu@post.kek.jp}}
        
\author{Massimo D'Elia\\
        Dipartimento di Fisica, Universit\`a di Pisa and INFN, Largo P
ontecorvo 2, I-56127 Pisa, Italy\\
        E-mail: \email{delia@df.unipi.it}}
                
\author{\speaker{Marco Mariti}\\
        Dipartimento di Fisica, Universi\`a di Pisa and INFN, Largo P
ontecorvo 2, I-56127 Pisa, Italy\\
        E-mail: \email{mariti@df.unipi.itl}}

\author{Francesco Negro\\
        Dipartimento di Fisica, Universit\`a di Genova and INFN,
Via Dodecaneso 33, I-16146 Genova, Italy\\
        E-mail: \email{fnegro@ge.infn.it}}

\abstract{We discuss our study of QCD in the 
presence of CP-odd electromagnetic (e.m.) 
background fields. We investigate the propagation 
of the CP-odd term from the e.m. sector to the strong 
sector, inducing an effective $\theta$ term. We discuss 
the method we have used in our lattice QCD simulations, 
and the results of our analysis, which are relevant to the 
determination of the effective pseudoscalar QED-QCD interactions. 
We also explore how these CP-odd e.m. background fields 
influence the number of the Dirac zero modes in our configurations.}

\FullConference{31st International Symposium on Lattice Field Theory - LATTICE 2013\\
		July 29 - August 3, 2013\\
		Mainz, Germany}

\begin{document}

\section{Introduction}
Theoretically one can add to the Euclidean action of QCD the additional term 
$- i\, \theta\, Q$, where
\beq
Q = \int d^4 x\, q(x) = 
\int d^4 x \frac{g^2}{64\pi^2} G_{\mu\nu}^a(x)
\tilde G_{\mu\nu}^a(x)
\label{thetaterm}
\eeq
is the topological charge operator, we defined
$G_{\mu\nu}^a$ as the non-Abelian gauge field strength and
$\tilde G_{\mu\nu}^a = 
\epsilon_{\mu\nu\rho\sigma} G_{\rho\sigma}^a$.

However, this term violates explicitly $CP$, while experiments tell us that QCD is invariant under $P$ and $CP$,
setting a quite stringent  upper bound on the parameter $\theta$,
which is expected to be
$|\theta| \lesssim 10^{-10}$~\cite{thetabound1,thetabound2}. Nevertheless, $\theta$ represents an important 
parameter in strong interactions, both from the theoretical and phenomenological point
of view.

Recently, it was hypothesized that 
local fluctuations of the topological charge
may induce measurable phenomena in the presence of
extremely intense magnetic fields. This scenario can be realized 
in non-central heavy ion collisions, where one expects
magnetic fields up to $10^{15}$ Tesla at LHC. According to the chiral magnetic 
effect~\cite{cme0,cme1}, for a magnetic field  strong enough to align the magnetic moments of quarks,
these local fluctuations of the topological charge would
induce a net unbalance of chirality, leading to a separation 
of electric charge along the direction of the magnetic field.

Albeit e.m. background fields couple directly only with charged particles,
recent lattice studies with dynamical fermions have shown that
these fields, via quark loop effects, can influence also the gluonic sector~\cite{lat1,lat2,lat3,lat4,lat5,latrev}. 
In an attempt to better clarify such issue, we investigate how the explicit breaking of the CP symmetry in the 
electromagnetic sector propagates to the gluon fields.

We will consider QCD in the presence of constant and uniform
electromagnetic background, such that 
$F_{\mu\nu} \tilde F_{\mu\nu} \propto \esb \neq 0$, which is expected to induce 
an effective CP-violating interaction in the non-abelian sector, 
$\tef \frac{g^2}{64\pi^2} G_{\mu\nu}^a(x) \tilde G_{\mu\nu}^a(x)$,
where $\tef$ must be an odd function of $\esb$. At the lowest order,
we can write
\beq
\tef \simeq \ceb e^2 \esb + O( (\esb)^3 )
\label{deftef}
\eeq
 where
$\ceb$ can be seen  as the susceptibility 
of the QCD vacuum to CP-breaking e.m. fields, and 
is directly related to the strength of the 
effective pseudoscalar QED-QCD interaction,
$\ceb\, q(x)\, e^2\, \esb$~\cite{elze1,mueller}.

In \cite{eb} we measured $\ceb$ on the lattice performing lattice
simulation of QCD in the presence of  e.m. background fields such
that $\esb \neq 0$, and we have determined the
induced $\tef$ by studying the distribution of the topological charge.
In this work we also report our study of the Dirac zero modes 
in the presence of such CP-odd e.m. background fields. 
 
\section{The method}

We can introduce the external e.m. fields in the QCD Lagrangean by modifying 
the quarks covariant derivative,
$D_\mu = 
\partial_\mu + i\, g A^a_\mu T^a + i\, q A_\mu$, where  $q$ 
is the quark electric charge and
$A_\mu$ is the e.m. gauge potential. That can be implemented on the lattice by adding 
appropriate $U(1)$ phases $u_\mu(n)$ to the usual $SU(3)$ parallel transports, i.e. making the substitution:
$U_\mu(n) \to u_\mu(n) U_\mu(n)$, where $n$ is a lattice site.
Because of the periodic boundary conditions used in
our lattice simulations, the possible value of the e.m. fields must be integer 
multiple of a minimum quantum:
\beq
f 
= {2 \pi}/({ q a^{2} L_\mu L_\nu})\, , 
\label{fquant}
\eeq
where $L_\mu$, $L_\nu$ are the lattice extensions along the 
directions orthogonal to the field (for more detail
see \cite{eb}). We have considered two flavour
QCD with dynamical fermions, using standard charges for the 
$up$ and $down$ quarks, namely
$q_u = 2 |e|/3$ and $q_d = - |e|/3$, therefore the quantization is given
in units of $f = {6 \pi}/({ |e| a^{2} L_\mu L_\nu})$.

To guarantee the feasibility of numerical simulations,
we must preserve the positivity of the fermion determinant after the
addition of the $U(1)$ phases to the $SU(3)$ link variables. This require
that the spectrum of the Dirac matrix in the path integral remains
purely imaginary.

However, such condition is not verified if we try to 
introduce a real electric field in Minkowski space:
it is easy to verify that this would require an
imaginary value of the electric field in Euclidean space,
which takes the $u_\mu$ variables 
out of the $U(1)$ group, making the fermion determinant complex: this
sign problem would hinder numerical simulations.

To circumvent this
problem we adopt the following strategy, used 
also in lattice studies of the electric 
polarizability of hadrons~\cite{shintani,alexandru}: we simulate real magnetic fields $\vec B$ and imaginary 
electric fields $\vec E  = i\, \vec E_I$ in Minkowski space, and then 
exploit analytic continuation. As a consequence, we expect to produce a purely imaginary effective
parameter $\tef = i\, \tief$. 

The presence of an imaginary $\theta_I$ adds, in the path integral, a factor
$\exp( \theta_I Q)$ to the probability 
distribution of gauge fields . This factor will shift the 
distribution of 
the topological charge by an amount which,
at the 
linear order in $\theta_I$, is given by the topological 
susceptibility $\chi$ at $\theta_I = 0 $:
\beq
\avq_{\theta_I} \simeq V\, \chi\, \theta_I = \avqs_{\theta = 0}\, \theta_I \ ,
\eeq  
here $V$ is the spacetime volume. 
That gives us the opportunity of 
determining the effective $\tief$ produced by a given e.m. field 
as
\beq
\tief \simeq
\frac{\avq (\vec E_I, \vec B)}{\avqs_0} + O( (\eisb)^3 )
\label{deftief}
\eeq
where $\langle \cdot \rangle_0$ is defined as the average taken 
at zero e.m. field. In the region of small $\tief$, which 
is the one relevant to Eq.~(\ref{deftef}), we expect negligible
corrections to Eq.~(\ref{deftief}).

\begin{figure}[htbp]
\centering
\begin{minipage}[c]{.485\textwidth}
\includegraphics[width=1.0\textwidth]{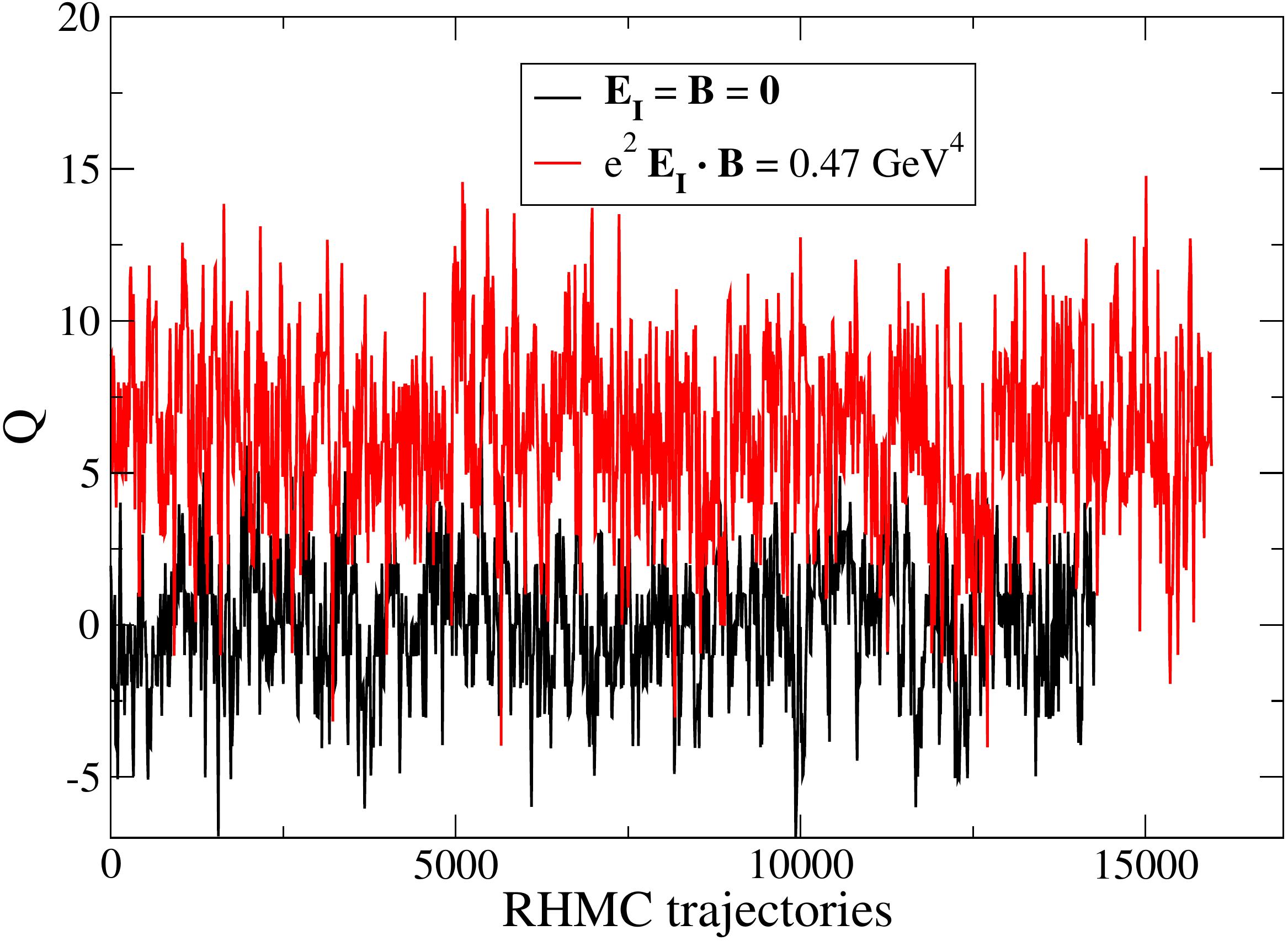}
\end{minipage}%
\hspace {3 mm}
\begin{minipage}[c]{.485\textwidth}
\includegraphics[width=1.0\textwidth]{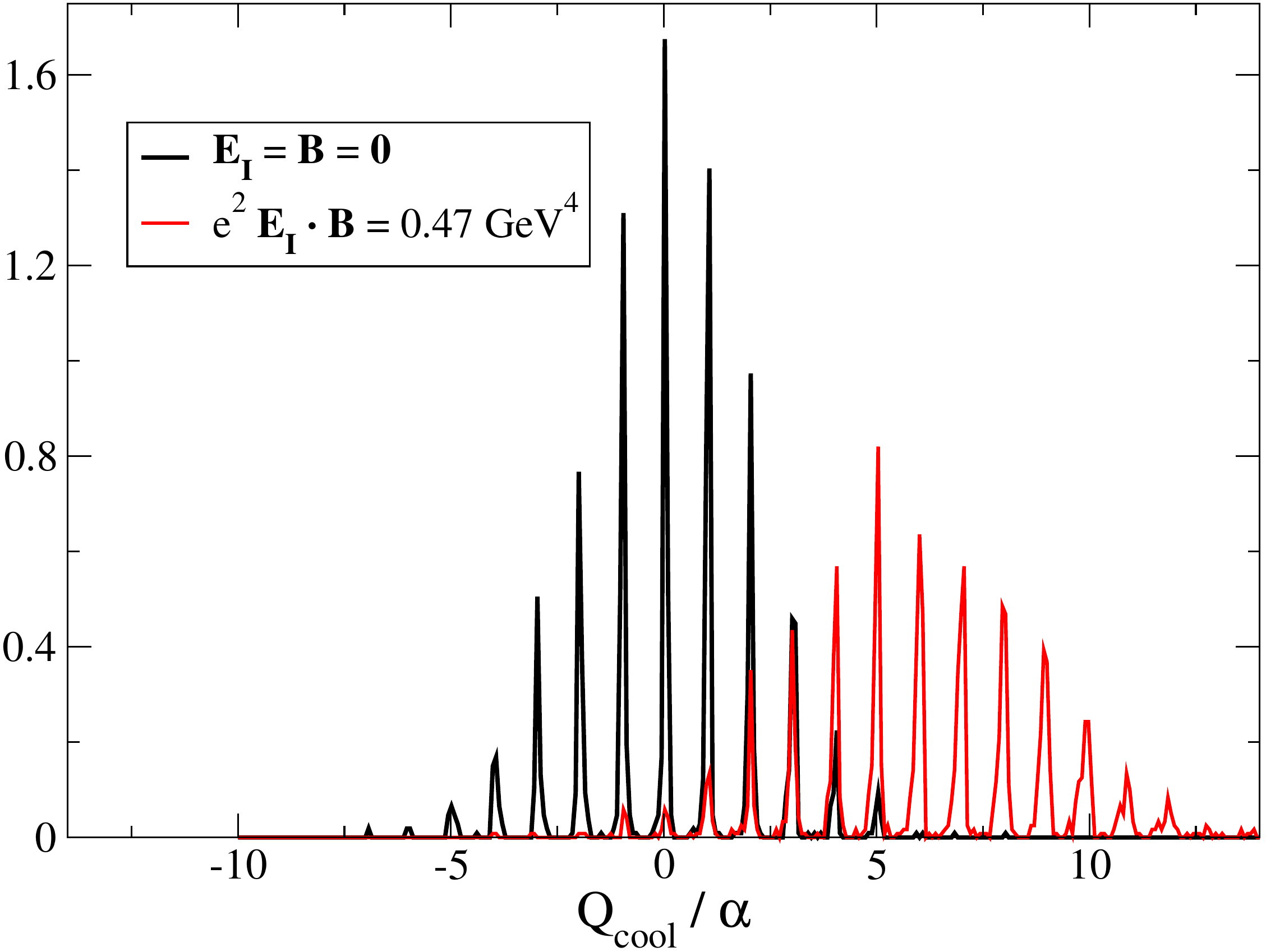}
\end{minipage}

\caption{Observed shift of the topological charge for 
$e^2\eisb = 0.47 \mbox{GeV}^4$ on a $16^4$ lattice, 
for $m_\pi\simeq~480$~MeV and $a\simeq0.113$ fm, 
for two values of the e.m. background  $\eisb$.
\emph{Left panel}: Monte Carlo histories.
\emph{Right panel}: topological charge distribution.}
\label{fig1}
\end{figure}

\section{Results} 

\begin{figure}[htbp]
\centering
\begin{minipage}[c]{.485\textwidth}
\includegraphics[width=1.0\textwidth]{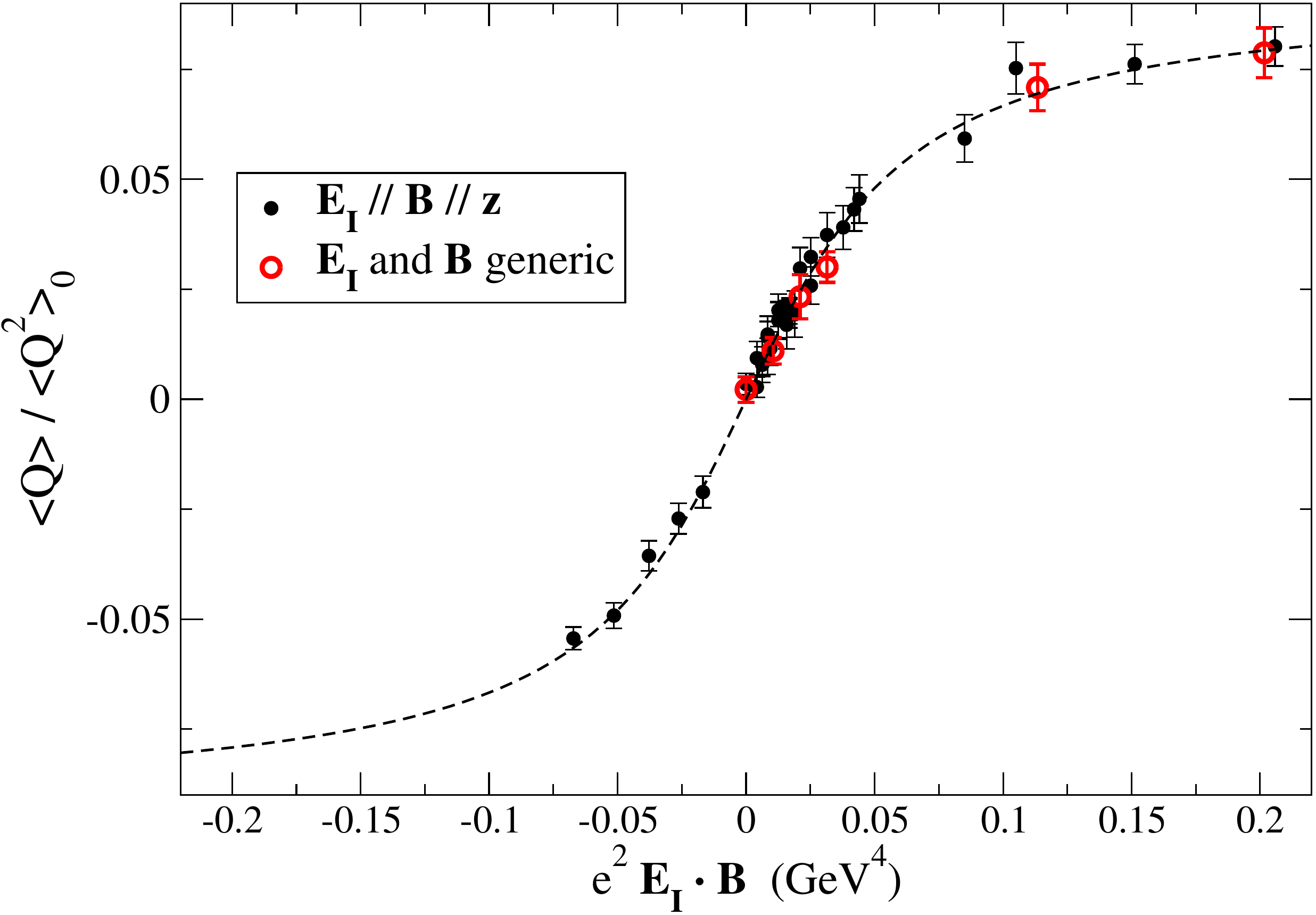}
\end{minipage}%
\hspace {3 mm}
\begin{minipage}[c]{.485\textwidth}
\includegraphics[width=1.0\textwidth]{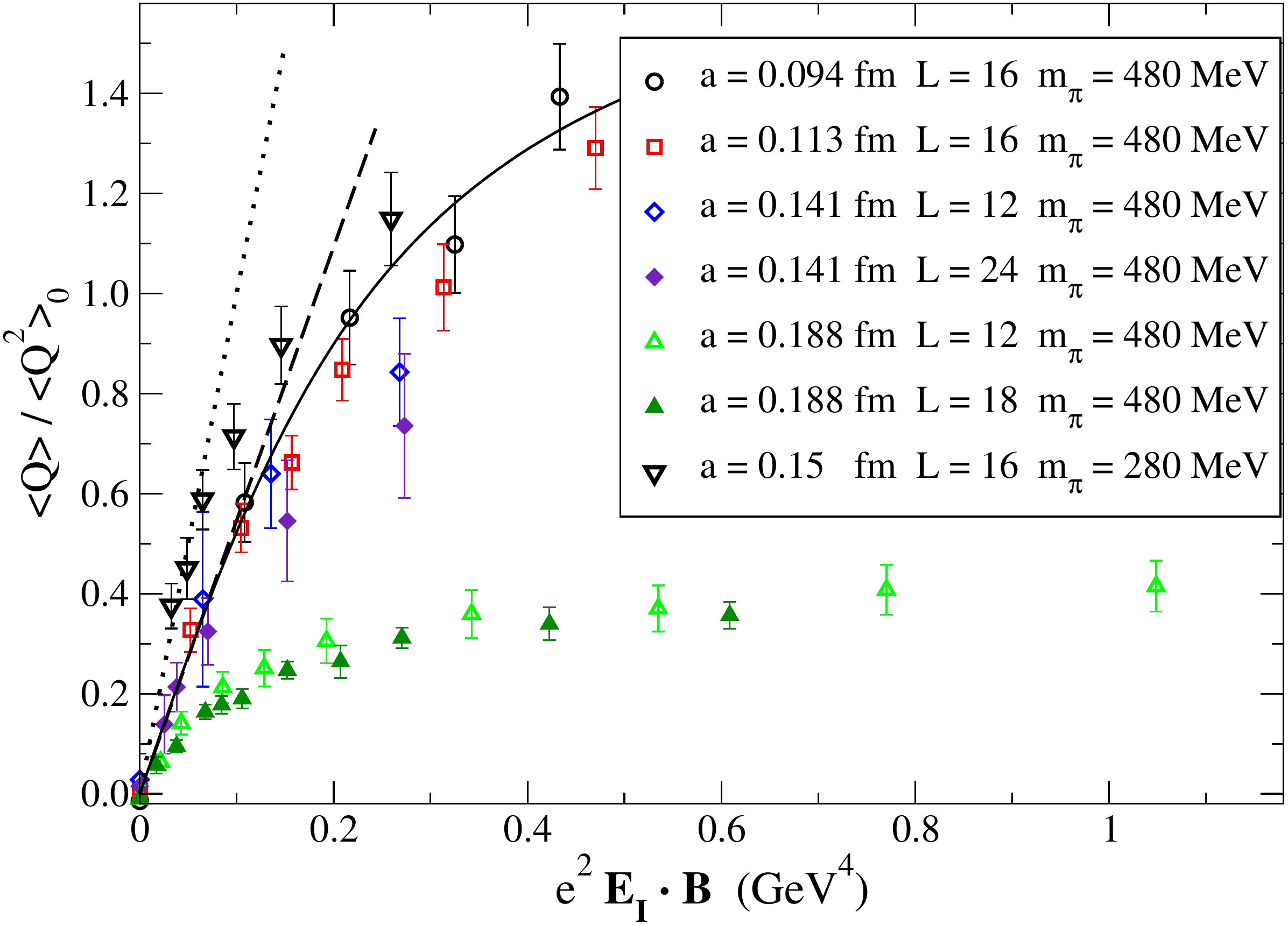}
\end{minipage}
\caption{\emph{Left panel}: ${\avq (\vec E_I, \vec B)}/{\avqs_0}$
for various ($\vec E_I,\,
\vec B$) on a $16^4$ lattice for $a \simeq 0.28$ fm and
$m_\pi \simeq 480$ MeV, where open circles 
corresponds to non parallel configurations of $\eisb/f^2$, while 
the dashed line is a best fit to Eq.~(\protect \ref{atanfun}). 
\emph{Right Panel}: ${\avq (\vec E_I, \vec B)}/{\avqs_0}$
as a function of $\eisb$ for different
spacings $a$ and lattice volumes $L^4$, and 
$m_\pi \simeq 480$ MeV.
The continuous line corresponds to a best fit
to Eq.~(\protect \ref{atanfun}) at the smallest value of $a$,
the dashed line is the corresponding slope at $\eisb = 0$.
We also plot our preliminary results for $m_\pi \simeq 
280$ MeV.
}
\label{fig2}
\end{figure}

We performed simulation of QCD with
$N_f = 2$ at $T = 0$
for a fixed pseudo-Goldstone pion mass
$m_\pi \simeq 480$ MeV. Different
lattice spacings have been explored by tuning the inverse gauge coupling
$\beta$ and $a m$ as described in Ref.~\cite{eb}.
We also used different lattice volumes
to check for
finite size corrections (see Fig.~\ref{fig2}). 

For the determination of $Q$ on gauge configurations, we adopted  
the standard discretized gluonic definition, measured after cooling~\cite{vicari_rep}, i.e.
recursive minimization of the 
pure gauge action to reduce ultraviolet (UV) artifacts.
We then rescaled the charge by a constant factor, so that its distribution gets peaked 
around an integer values, (see, e.g., Fig \ref{fig1}), and we finally fix $Q$ to the closest
integer (for details and discussions on the used procedure see~\cite{eb}).

In the left panel of Fig.~\ref{fig1} we show 
the Monte-Carlo history of the topological charge for two 
numerical simulations performed respectively
at $\eisb~=~0$ and 
$e^2~\eisb~\simeq~0.47\,~{\rm GeV}^4$.
We see that, when we switch on the external fields $\eisb\neq0$, 
the fluctuations of $Q$ shifts from zero 
toward positive values, as clearly appears also
from the right panel of Fig.~\ref{fig1}, where we plot the corresponding distributions of $Q$.

To  better investigate the effective dependence 
of $\avq (\vec E_I, \vec B)$ on the background field combination $\eisb$,
in Fig.~\ref{fig2} we show
${\avq (\vec E_I, \vec B)}/{\avqs_0}$, where data are obtained for a variety
 of combinations of $\vec E_I$ and 
$\vec B$, mostly taken parallel to the $z$ axis,
and then plotted versus $\eisb$. The fact that all
data fall on the same curve, even when $\vec E_I$ and $\vec B$
are not parallel, is a nice demonstration that 
$\tief$ is indeed a function of $\eisb$ alone, as expected. We have
also considered different combinations
of the fields having the same or opposite values for $\eisb$, to verify
 explicitly that $\tief$ is odd in $\eisb$. 

 For small fields we observe a linear dependence in $\eisb$, 
 while for larger fields the observable shows saturation effects, as is common to many
systems having a linear response to external stimulation.
All data can be nicely fitted by the function
\beq
{\avq (\vec E_I, \vec B)}/{\avqs_0} = 
a_0\ {\rm atan} (a_1 \esb ) \ ,
\label{atanfun}
\eeq
the best fit curve is shown
in Fig.~\ref{fig2}, corresponding to $\chi^2/{\rm d.o.f.} = 0.74$.

We expect ${\avq (\vec E_I, \vec B)}/{\avqs_0}$ to be $V$ independent,
because $\avqs_0$ and $\avq$ are both derivatives of the free energy with respect of $\theta$,
so they are proportional to $V$, and their ratio should be volume independent.
In Fig.~\ref{fig2} we show 
${\avq (\vec E_I, \vec B)}/{\avqs_0}$ 
for
$m_\pi \simeq 480$ MeV and 
different
spacings $a$ and lattice volumes $L^4$.
From the right panel in Fig.~\ref{fig2} we can exclude relevant finite size effects, 
even on the smallest volumes explored, corresponding
to $a m_\pi L \sim 4$.

Instead, we observed a significant dependence on the UV cutoff until
$a \lesssim 0.15$ fm. Apart from standard lattice artifacts related
to the path integral discretization, additional systematic effects
may be related to 
the method used to determine $Q$: if $a$
is coarse enough that part of the induced topological background 
lives close to the UV scale, 
then the cooling procedure is expected to destroy part of such background.
However data obtained for $a \lesssim 0.15$ fm are in 
very good agreement with each other, particularly in the region of small values of $\eisb$, 
where corrections to Eq. (\ref{deftief}) should be negligible.

We determined $\ceb$ performing best fits of the data
in Fig.~\ref{fig2} to the function in Eq.~(\ref{atanfun}), in a range of fields such that
$e^2 \eisb < 0.8$ GeV$^4$, then
considering its slope at $\eisb = 0$ and exploiting Eqs.~(\ref{deftef})
and (\ref{deftief}).
For each slope we obtained a good agreement with a direct linear fit
performed on a narrow region of small $\eisb$. 
Because of the large artifacts at coarse lattice spacing, 
we consider only data up to $a < 0.15$ to extrapolate our result, finding 
$\chi_{CP}=5.47(78)\ \mbox{GeV}^{-4}$ ($\chi^2/$dof $\simeq\ 0.1$).
We also expect an additional $\sim20\%$ uncertainty on $\chi_{CP}$ coming out
from a $5\%$ systematic uncertainty in our knowledge of $a$.  Preliminary results
obtained on a $16^4$ lattice and for $a \simeq 0.15$ fm 
indicate instead 
$\ceb \sim 10$ GeV$^{-4}$ if $m_\pi \simeq 280$ MeV, 
suggesting that $\ceb$ tends to increase when approaching the chiral
limit.

\section{Zero modes}
\begin{figure}[htbp]
\centering

\includegraphics[width=0.5\textwidth]{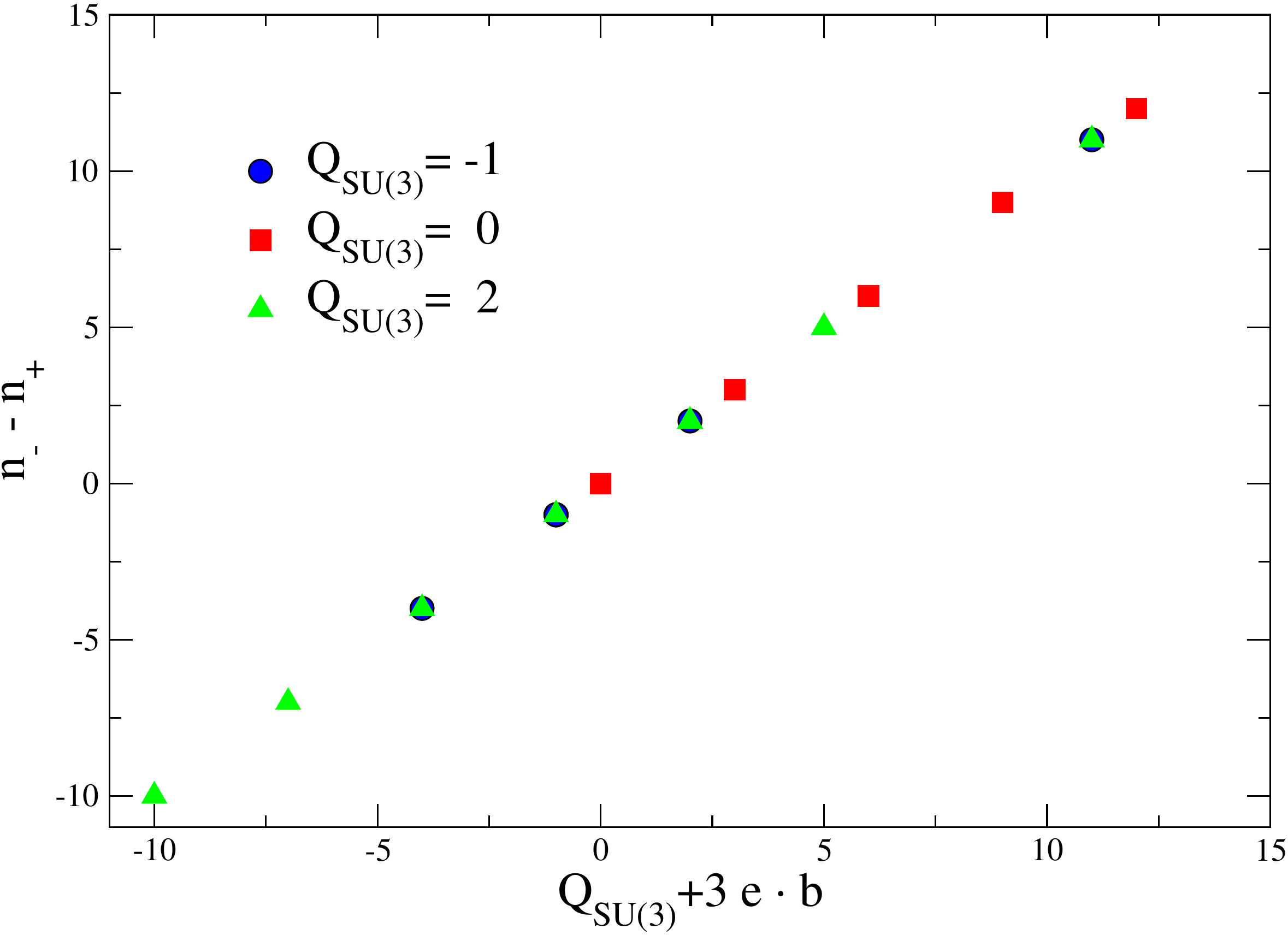}

\caption{$n_{-} - n_{+}$ for three different values of the topological 
charge $Q_{SU(3)}$ and for various combination of the electric 
and magnetic fields (here given in units of their quanta).}
\label{fig3}
\end{figure}
From our lattice results it clearly  appears that CP-odd e.m. background 
have a non trivial influence on the 
gluon fields, shifting the total distribution of the topological charge to finite values.
From a naive consideration, these configurations should be
suppressed in the path integral, because of their higher value of the action.
Moreover, the axial anomaly equation tell us that a non zero
value of  $Q$ is associated with the presence of
zero modes in the fermion matrix, which should drop in the chiral limit, 
the contribution of these configurations.

To explain the observed phenomena (and the apparent increase of
its strength for smaller masses) one has to consider the full axial anomaly
equation, with the inclusion of the $U(1)$ term
brought from the external e.m. fields, $Q_{U(1)}=\esb$. Then the full axial anomaly 
equation become:
\beq
n_{-} - n_{+} = Q_{tot} \equiv Q_{SU(3)} + N_C\ Q_{U(1)}
\label{zero}
\eeq
where $Q_{SU(3)}$ and $Q_{U(1)}$ are respectively the non-abelian
and the abelian contributions to $Q_{tot}$, which is the 
difference of left handed and right handed 
zero modes. 

To verify explicitly this relation, 
we measured  the number of zero modes
in a set of $O(40)$ configurations, where we have fixed both the 
topological content and the
external fields. To measure 
the zero modes we used overlap fermions, which, as well known,
can correctly distinguish the chirality of fermions on the lattice.
We explicitly verified the relation~(\ref{zero}), with $N_C = 3$, for various external e.m. 
fields and three different values of the topological charge $Q_{SU(3)}$ (see Fig.~\ref{fig3}).
So, at least in the chiral limit, the relevant gluonic configurations in the path integral
must have nontrivial $Q_{SU(3)}$ in such a way to balance the contributions
carried by the electromagnetic part of the anomaly.

\section{Discussion}

Our results can be compared with the
phenomenological estimate given in 
Ref.~\cite{mueller}, where the authors based their calculations
on the effective couplings of the $\eta'$ and $\eta$ mesons 
to two gluons
and to two photons, finding $\ceb \approx 0.73/(\pi^2 f_\eta^2 m_{\eta'}^2) \sim
3$ GeV$^{-4}$.  Our measurements suggest that the lattice QCD result 
for the effective pseudoscalar QED-QCD interaction is larger,
even if of the same order of magnitude. However, one has to 
consider the different systematics 
(the phenomenological estimate is based on a theory with 2+1 light flavors),
and the unphysical value of the quark mass used in our simulations.

Concerning the validity of analytic continuation from imaginary to 
real electric fields in Min-kowski space, 
a real, non-zero and constant electric field, even if infinitesimal,
will induce vacuum instabilities in the thermodynamical
limit . On the other hand 
this should not be true in presence of an infrared cutoff, i.e. if 
electric fields are limited in space. Therefore 
our result should be useful for the determination of a local
effective $\theta$ parameter produced by 
smooth and limited in space CP-odd
e.m. fields. It would be interesting
in the future to repeat this analysis with smoothly 
varying fields, as well as with physical quark masses
and at finite temperature.

\section{Acknowledgments}
We thank Philippe de Forcrand for useful discussions.
Numerical simulations have been carried out on two GPU
farms at INFN-Pisa and INFN-Genoa and on the
QUONG GPU cluster in Rome.

\end{document}